\documentclass[a4paper]{jpconf}
\usepackage{graphicx}
\usepackage{amsmath}
\usepackage{amssymb}

\newcommand{\sNN}{s_\mathrm{NN}}

\begin{document}
\title{Estimating $\eta/s$ of QCD matter at high baryon densities}

\author{Iu.~Karpenko$^{1,2}$, M. Bleicher$^{1,3}$, P. Huovinen$^{1,3}$ and H. Petersen$^{1,3,4}$}

\address{$^1$ Frankfurt Institute for Advanced Studies, Ruth-Moufang-Stra{\ss}e 1, 60438 Frankfurt am Main, Germany}
\address{$^2$ Bogolyubov Institute for Theoretical Physics, 14-b, Metrolohichna st., 03680 Kiev, Ukraine}
\address{$^3$ Institute for Theoretical Physics, Johann Wolfgang Goethe Universit\"at, Max-von-Laue-Str.~1, 60438 Frankfurt am Main, Germany}
\address{$^4$ GSI Helmholtzzentrum für Schwerionenforschung GmbH, Planckstr. 1,  64291 Darmstadt, Germany}
\ead{karpenko@fias.uni-frankfurt.de}

\begin{abstract}
We report on the application of a cascade + viscous hydro + cascade model for heavy ion collisions in the RHIC Beam Energy Scan range, $\sqrt{\sNN}=6.3\dots200$~GeV. By constraining model parameters to reproduce the data we find that the effective(average) value of the shear viscosity over entropy density ratio $\eta/s$ decreases from 0.2 to 0.08 when collision energy grows from $\sqrt{\sNN}\approx7$ to 39~GeV.
\end{abstract}

\section{Introduction}

The principal goal of the Beam Energy Scan program at Relativistic Heavy Ion Collider (RHIC) is to study the transition from baryon-free to baryon-rich regime of strongly interacting QCD matter by decreasing the collision energy. One of questions is what the transport properties of the medium with high baryon density, created at lower beam energies, are.

To address this question we use a state-of-the-art event-by-event viscous hydrodynamic + cascade (or hybrid) model to describe the evolution of the collision system.

\section{The model, its parameter space and results}

The hybrid model consists of the pre-equilibrium stage, the fluid dynamical stage and the final hadronic cascade. The pre-thermal stage of evolution is described by UrQMD cascade \cite{Bass:1998ca, Bleicher:1999xi}. The interactions are calculated until a hypersurface of a constant Bjorken proper time $\tau=\sqrt{t^2-z^2}=\tau_0$, which is a parameter of the model. At $\tau=\tau_0$ the energy, momentum and charges of particles are smoothly distributed to fluid cells according to a Gaussian profile. The contribution of each particle to a fluid cell $\{i,j,k\}$ is given by:
\begin{align}
\left\{ \Delta P^\mu_{ijk}, \Delta N^0_{ijk} \right\}&=\left\{ P^\mu, N^0 \right\} \cdot C\cdot\exp\left(-\frac{\Delta x_i^2+\Delta y_j^2}{R_\perp^2}-\frac{\Delta\eta_k^2}{R_\eta^2}\gamma_\eta^2 \tau_0^2\right). 
\end{align}
where $\Delta x_i$, $\Delta y_j$, $\Delta \eta_k$ are the differences
between particle's position and the coordinates of the hydrodynamic
cell $\{i,j,k\}$.
Due to the lack of knowledge about thermalization process, $R_\perp$ and $R_\eta$ are another free parameters of the model.
These distributions are used as an initial state for 3-dimensional hydrodynamic expansion, described by relativistic viscous hydrodynamic code \texttt{vHLLE} \cite{Karpenko:2013wva}. We use the equation of state based on Chiral model \cite{Steinheimer:2010ib} in the fluid stage. We set bulk viscosity and baryon charge diffusion to zero in the fluid phase, and study the effects of shear viscosity only. When the energy density reaches a certain value $\epsilon=\epsilon_\text{sw}$, the hydrodynamic medium is converted to particles (particlized) according to Cooper-Frye prescription \cite{Cooper:1974mv}. The last stage of evolution - hadron rescatterings and  decays - is described again with UrQMD cascade. For more details see Ref. \cite{Karpenko:2015xea}.

\begin{figure}
\includegraphics[width=0.5\textwidth]{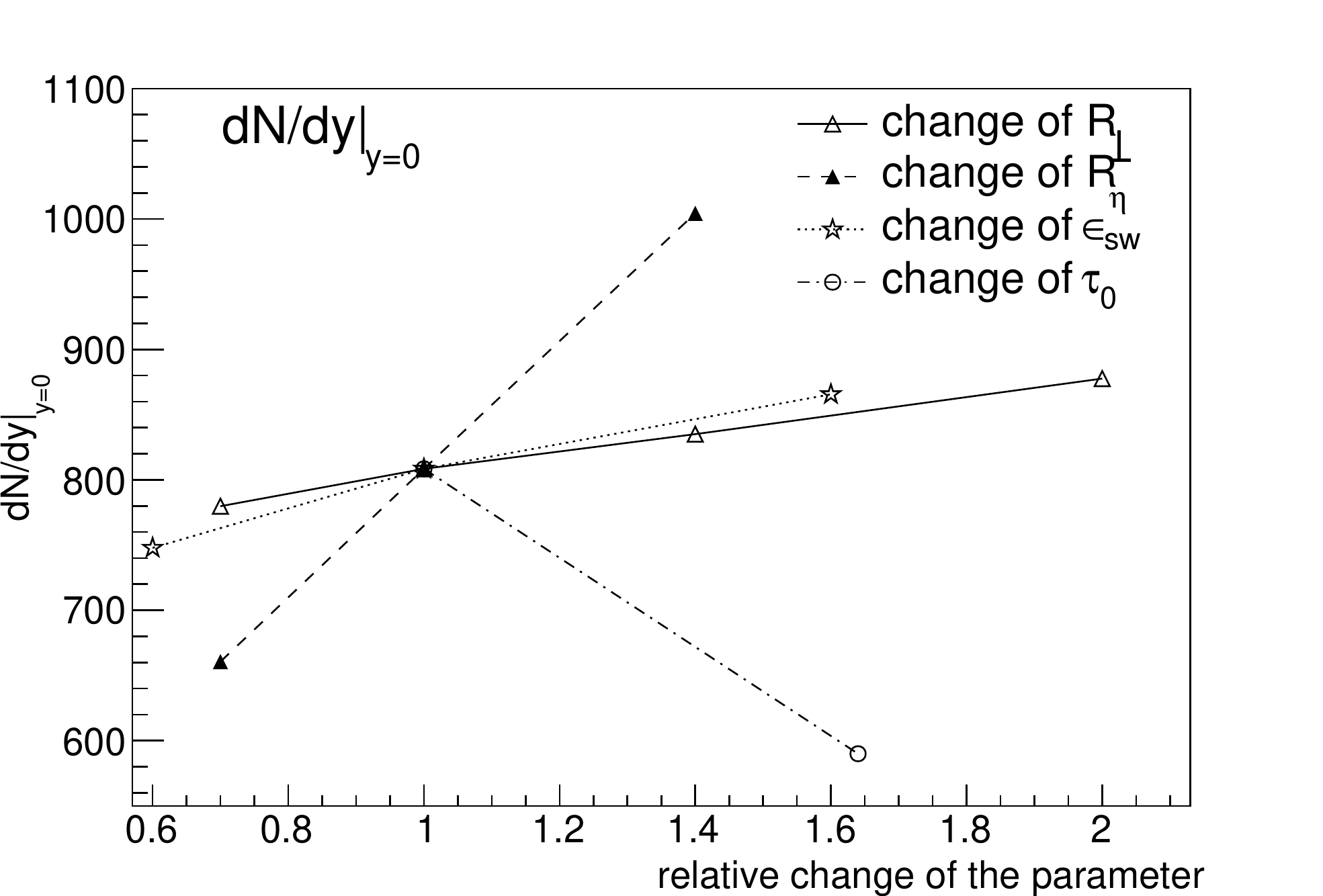}
\includegraphics[width=0.5\textwidth]{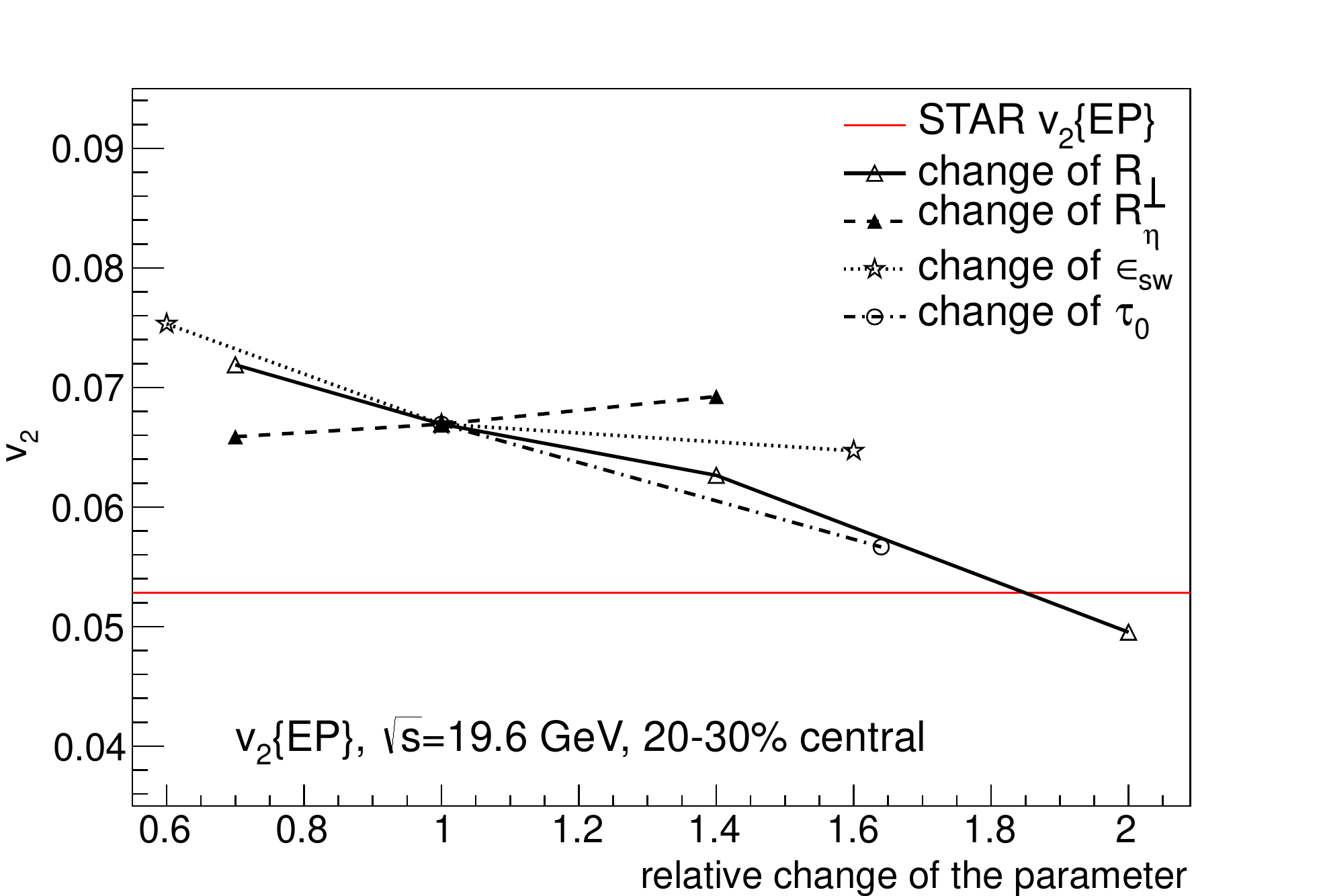}
\caption{Parameter dependence of the total yield at midrapidity (left) and $p_T$ integrated elliptic flow (right) in 20-30\% central Au-Au collisions at $\sqrt{\sNN}=19.6$~GeV. Figures are from Ref.~\cite{Karpenko:2015xea}.}\label{figParamDep}
\end{figure}

As a result, the model has three parameters related to the initial state (starting time $\tau_0$ and Gaussian smearing radii $R_\perp$, $R_\eta$), one parameter of the hydrodynamic medium (shear viscosity over entropy ratio $\eta/s$) and one parameter related to switching from hydrodynamics to cascade (switching energy density $\epsilon_\text{sw}$). The main goal is to extract the value of the \textit{physical} parameter $\eta/s$ from comparison with available experimental data. This goal is, however, complicated by the fact that the observables in the model depend significantly on all the mentioned parameters, not only $\eta/s$. A somewhat stronger sensitivity to the initial state parameters compared to hydrodynamic studies at full RHIC or LHC energies [someone] is explained by shorter duration of hydrodynamic phase at lower beam energies.

\begin{figure}
\includegraphics[width=0.5\textwidth]{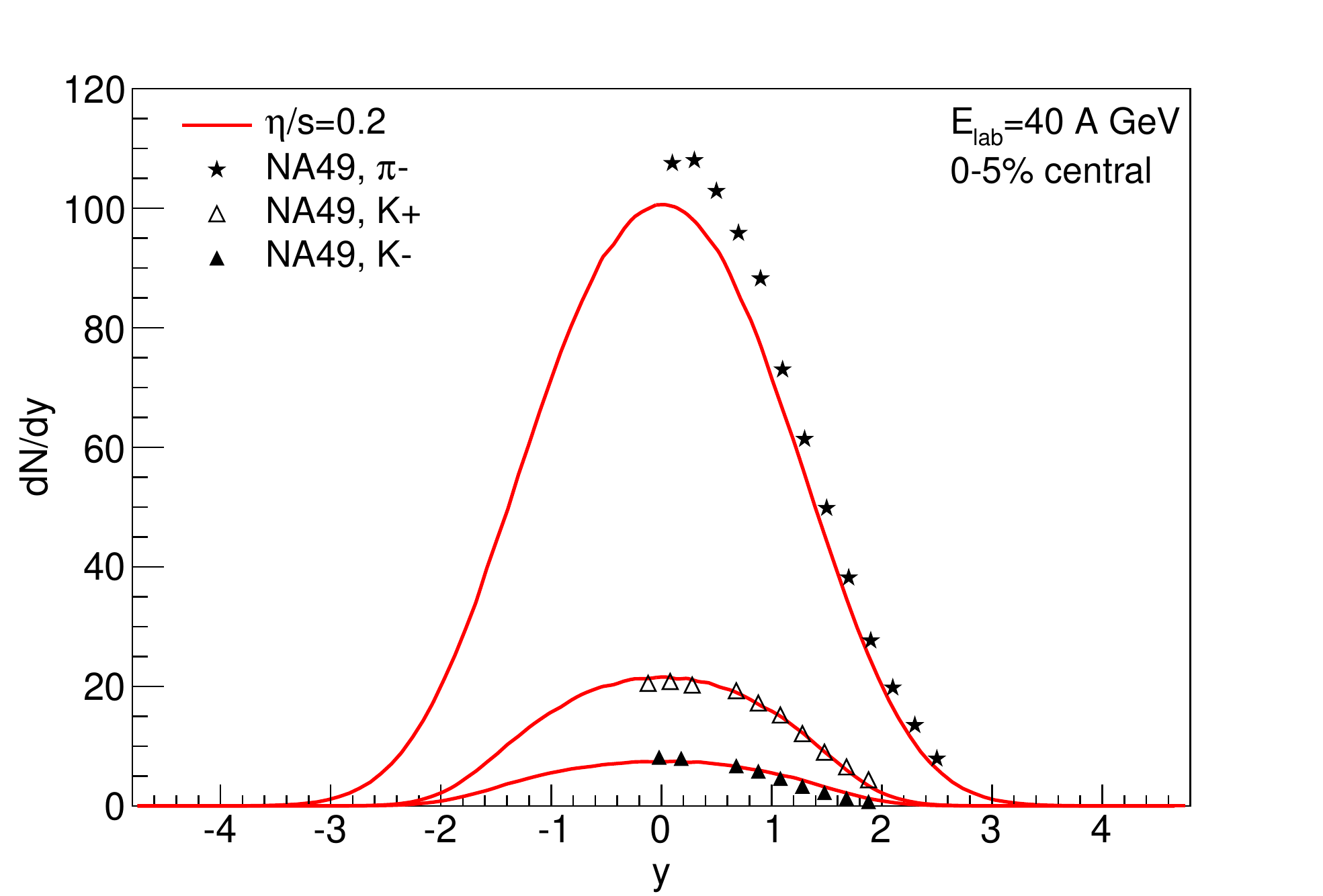}
\includegraphics[width=0.5\textwidth]{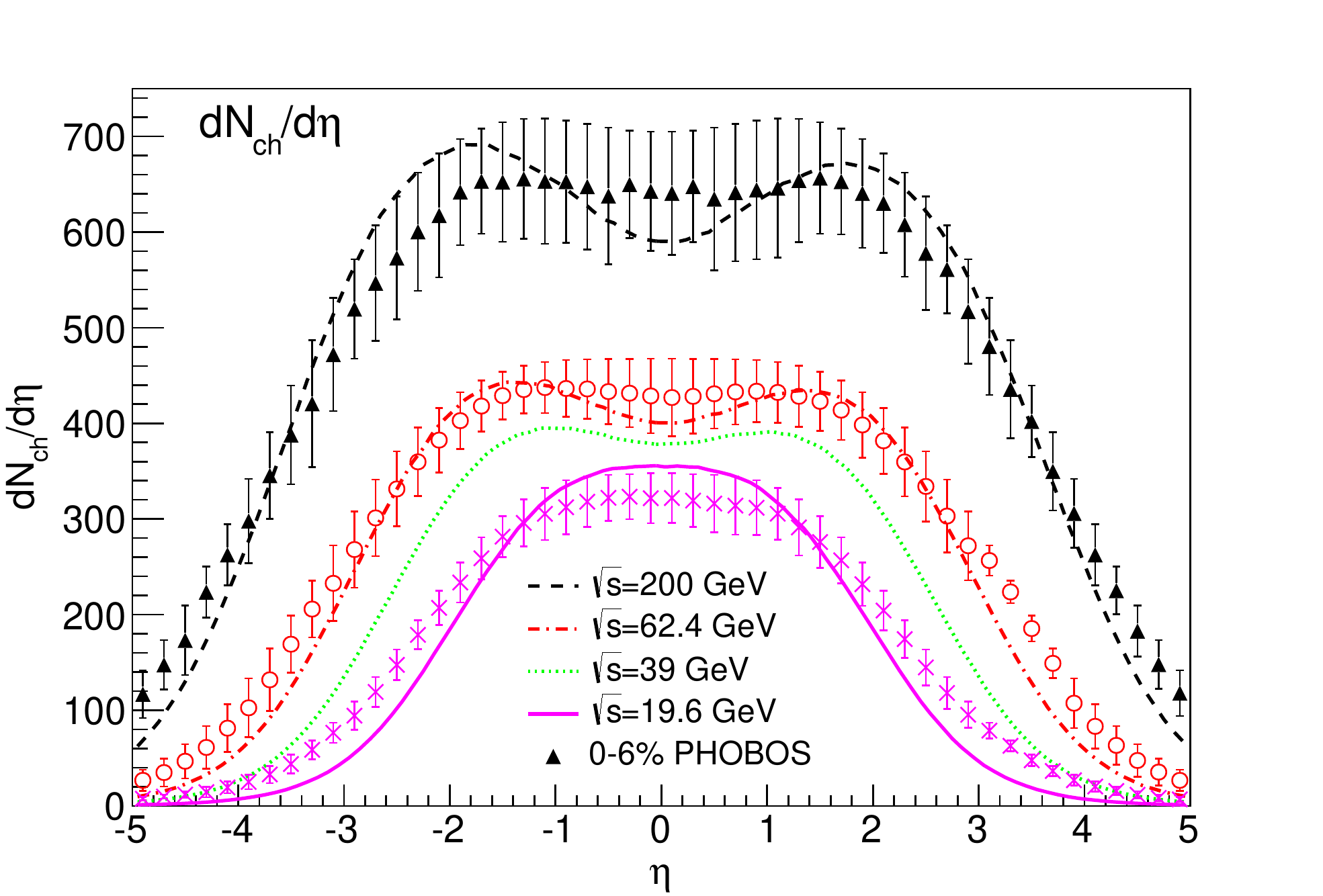}\\
\includegraphics[width=0.5\textwidth]{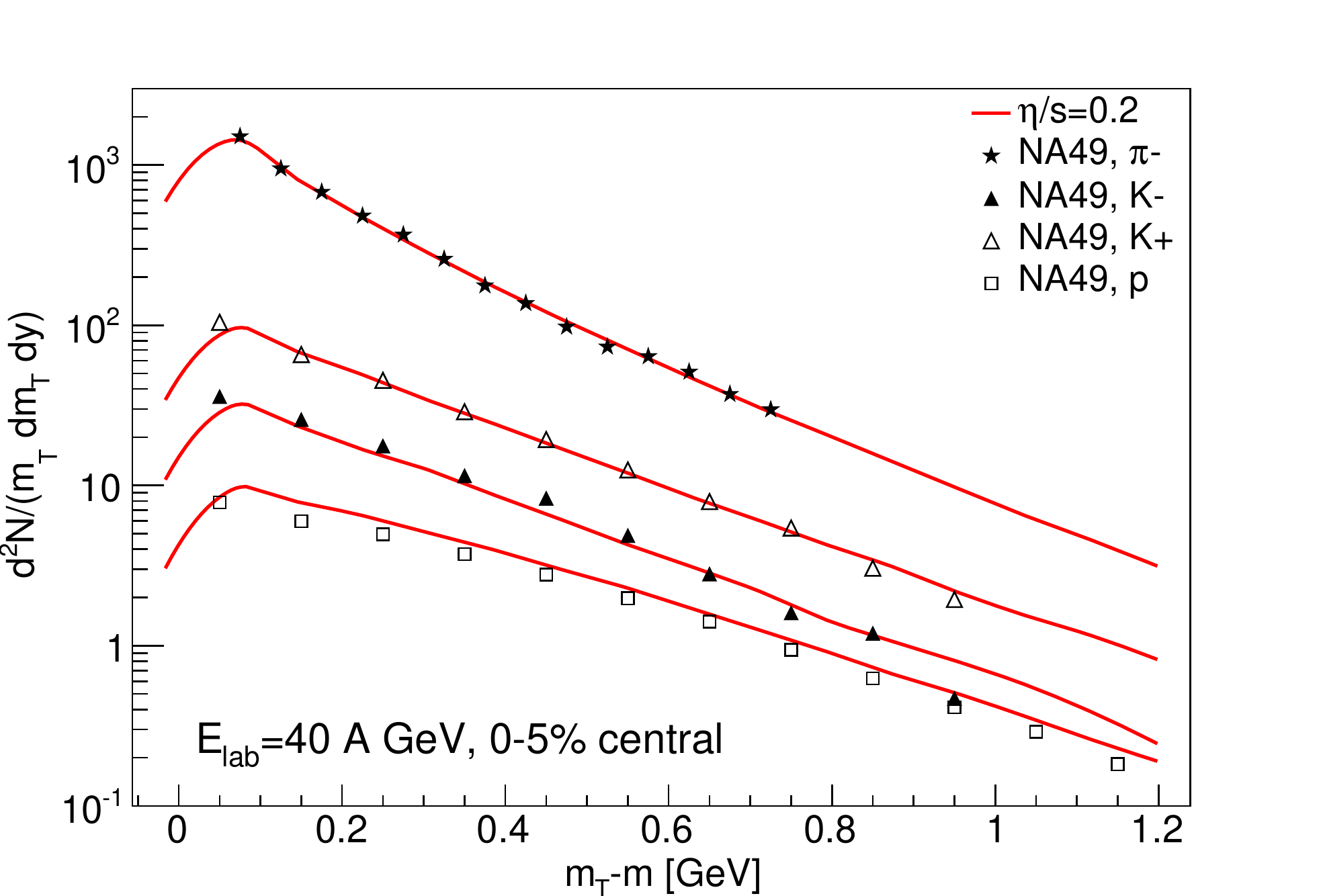}
\includegraphics[width=0.5\textwidth]{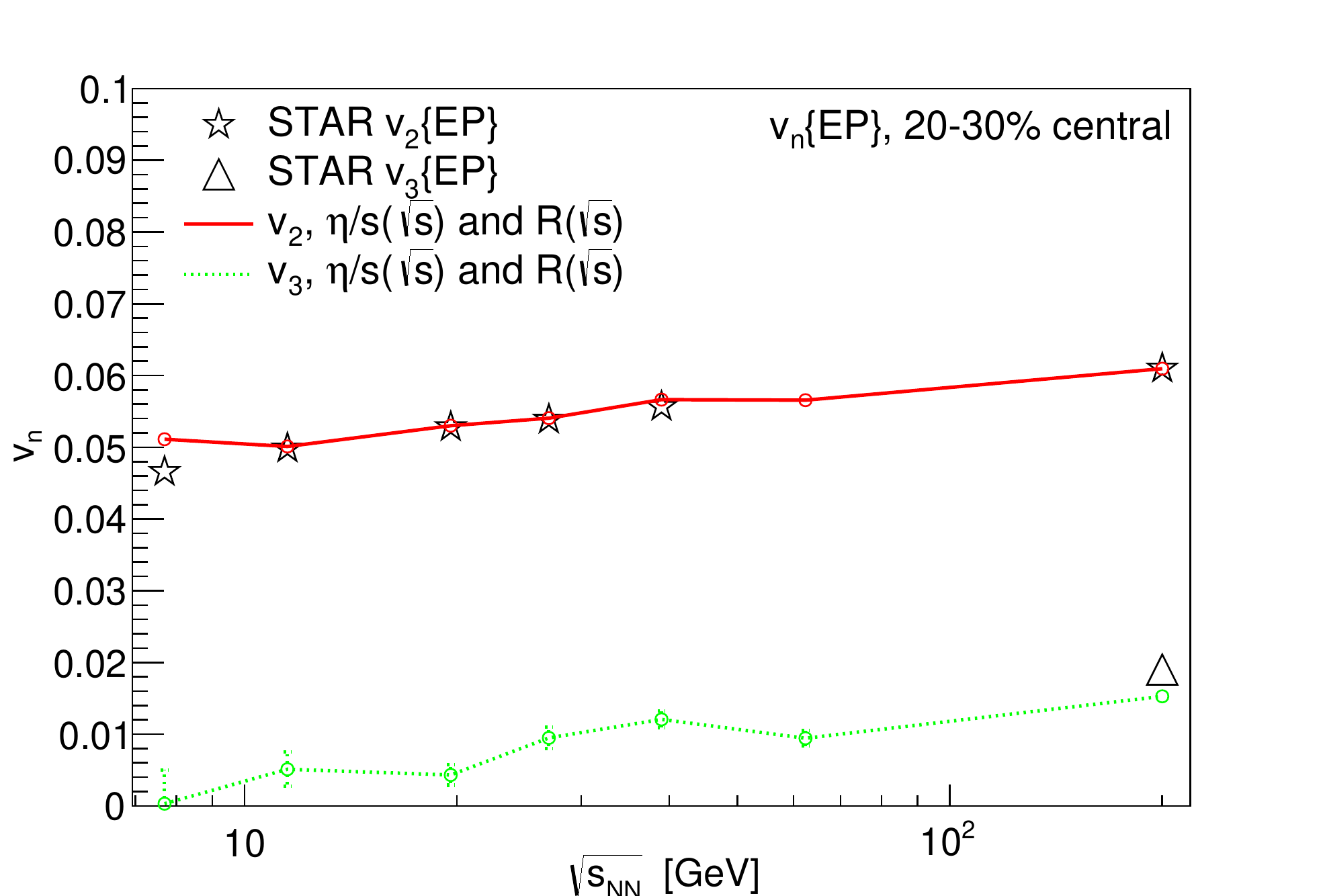}\\
\vspace{-10pt}
\caption{Rapidity (top left) and transverse momentum (bottom left) distributions of pions and kaons at $E_{\rm lab}=40$~A~GeV Pb-Pb collisions, pseudorapidity distributions of all charged hadrons at $\sqrt{s}=19.6,62.4$~and~$200$~GeV Au-Au collisions (top right), and $p_T$ integrated flow coefficients for different collision energies \cite{Adamczyk:2012ku} (bottom right), all calculated in the model and compared to the experimental data from NA49 \cite{Afanasiev:2002mx} and PHOBOS \cite{Alver:2010ck}. Figures are from Ref.~\cite{Karpenko:2015xea}.}\label{figResults}
\vspace{-10pt}
\end{figure}

To learn how exactly final observables depend on the model parameters, we consider the setup with $R_{\perp}=R_\eta=1.0$~fm, $\tau_0=2R/(\gamma v_z)$, $\eta/s=0$ and $\epsilon_{\rm sw}=0.5$~GeV/fm$^3$ as a ``default'' scenario. On Fig.~\ref{figParamDep} we show how deviations of parameter values from the ``default'' scenario affect the multiplicity of all hadrons at midrapidity and  $p_T$ integrated elliptic flow of all charged hadrons at one particular collision energy $\sqrt{\sNN}=39$~GeV. One can summarize the dependencies seen in Fig.~\ref{figParamDep} in a following way:
\begin{itemize}
 \item Increased $R_\perp$ smoothens the initial energy density
   profile in the transverse plane, which leads to smaller gradients
   and less explosive transverse expansion. Increased $R_\perp$ also results in decreased ellipticity an initial energy density profile, which is hydrodynamically translated into smaller final elliptic flow.
 \item In a similar manner, the increase of $R_\eta$ leads to shallower
   longitudinal gradients and weaker longitudinal expansion. Thus
   more energy remains at midrapidity to form stronger transverse expansion,
   which increases $v_2$. On the other hand, larger $R_\eta$ also results in larger initial entropy of the fluid, which considerably increases the final particle multiplicity.
 \item Increased $\tau_0$ leads to a shorter lifetime of the hydrodynamic phase, as a result of longer pre-thermal phase. At the same time $\tau_0$ enters the Gaussian energy/momentum smearing profile. Thus its increase acts opposite to the increase of $R_\eta$. The results show that the second effect is stronger.
 \item Increased $\epsilon_{\rm sw}$ shortens the effective lifetime of the hydrodynamic
   phase. The shorter time to develop radial and elliptic flows 
   is not fully compensated by the longer cascade phase,
   which results in smaller final $v_2$.
   Since the total entropy is conserved in the ideal hydrodynamic expansion, but increases in the cascade stage, the final particle multiplicity increases with the increase of $\epsilon_{\rm sw}$.
\end{itemize}
The found dependencies allow us to adjust the values of parameters individually for every collision energy to approximately reproduce the available experimental data points in the BES region. To simplify the analysis, we keep the same value of switching energy density $\epsilon_\text{sw}$ at all collision energies. The adjusted values of the parameters are shown in Table~\ref{tbParameters}. Some of the results are shown in Fig.~\ref{figResults}.

The shear viscosity over entropy density ratio $\eta/s$ in fluid phase is taken to be independent of temperature and chemical potentials, and changes only with collision energy. Therefore the value of $\eta/s$ can be thought of as effective one for given computational setup. The parameter adjustment procedure yields $\eta/s$ values which decrease with increasing collision energy. At lower collision energies the fluid phase probes thermodynamic region of lower temperature, limited by the particlization temperature from below, and higher chemical potential. One expects that physical $\eta/s$ depends on temperature and chemical potential, and not on collision energy. Therefore we conclude that the $\sqrt{\sNN}$ dependence of the effective $\eta/s$ we found hints that in more realistic simulations $(\eta/s)(T,\mu_B)$ should increase with increasing baryon chemical potential $\mu_B$.

\begin{table}
\begin{tabular}{|l|l|l|l|l|}
\hline
 $\sqrt{\sNN}$~[GeV] & $\tau_0$~[fm/c] & $R_\perp$~[fm] & $R_\eta$~[fm] & $\eta/s$ \\ \hline
     7.7          &      3.2        &     1.4        &     0.5    &    0.2   \\ \hline
     8.8 (SPS)    &      2.83       &     1.4        &     0.5    &    0.2   \\ \hline
     11.5         &      2.1        &     1.4        &     0.5    &    0.2   \\ \hline
     17.3 (SPS)   &      1.42       &     1.4        &     0.5    &    0.15  \\ \hline
     19.6         &      1.22       &     1.4        &     0.5    &    0.15  \\ \hline
     27           &      1.0        &     1.2        &     0.5    &    0.12  \\ \hline
     39           &      0.9*        &     1.0        &     0.7    &    0.08  \\ \hline
     62.4         &      0.7*        &     1.0        &     0.7    &    0.08  \\ \hline
     200          &      0.4*        &     1.0        &     1.0    &    0.08  \\ \hline
 \end{tabular}
 \hspace{10pt}
\begin{minipage}[m]{15pc}\caption{
Collision energy dependence of the model parameters, which allows the model to reproduce the experimental data in BES region and higher RHIC energies.}\label{tbParameters}
\end{minipage}
\end{table}
\vspace{-10pt}

\section{Summary}

To summarize, the 3 dimensional cascade + viscous hydro + cascade model is applied for the description of heavy ion collisions in the collision energy range of RHIC Beam Energy Scan program, $\sqrt{\sNN}=6.3\dots200$~GeV. The parameter space of the model is explored. The values of model parameters are adjusted for different collision energies in order to approach the existing experimental data for rapidity and $p_T$ distributions, and elliptic flow coefficients of hadrons. As a result it is found that the value of effective shear viscosity over entropy density ratio $\eta/s$ decreases from 0.2 to 0.08 when the center of mass collision energy increases from 7.7 to 39 GeV, and stays at 0.08 for $\sqrt{\sNN}>39$~GeV.

\section*{Acknowledgements}
 The simulations have been performed at the Center for Scientific Computing (CSC) at the Goethe-University Frankfurt. The authors acknowledge the financial support by the Helmholtz International Center for FAIR
 and Hessian LOEWE initiative. HP acknowledges funding by the Helmholtz Young Investigator Group VH-NG-822.  The work of PH was supported by BMBF under contract no. 06FY9092.

\end{document}